\documentclass[aps,pre,twocolumn,nofootinbib]{revtex4}

\usepackage[final]{graphics}
\usepackage{amsmath,pstricks}
\usepackage{amsfonts,amsbsy}
\usepackage[latin1]{inputenc}  

\usepackage{graphicx}
\usepackage{amsmath,amssymb,amsthm}

\newcommand {\ow}{\overline{\omega}}
\newcommand {\og}{\overline{G}}
\newcommand {\fs}{\stackrel{.}{=}}

\def\k{{\boldsymbol k}}

\newcommand{\IR}{{\Bbb R}}

\begin{document}

\title{Zero-mode analysis of quantum statistical physics}

\author{A. {\sc Bessa}$^{1}$\footnote{abessa@if.usp.br},
C. A. A. {\sc de Carvalho}$^{2}$\footnote{aragao@if.ufrj.br}
and E. S. {\sc Fraga}\footnote{fraga@if.ufrj.br}}

\affiliation{$^{1}$Instituto de F\'\i sica, Universidade de S\~ao Paulo,
Caixa Postal 66318, 05315-970, S\~ao Paulo, SP , Brazil \\
$^{2}$Instituto de F\'\i sica, Universidade Federal do Rio de Janeiro, 
Caixa Postal 68528, 21941-972, Rio de Janeiro, RJ , Brazil}

\date{\today}

\begin{abstract}
We present a unified formulation for quantum statistical physics based on the representation of the 
density matrix as a functional integral. We identify the stochastic variable of the effective statistical
theory that we derive as a boundary configuration and a zero mode relevant to the discussion
of infrared physics. We illustrate our formulation by computing the partition function of an interacting 
one-dimensional quantum mechanical system at finite temperature from the path-integral representation for 
the density matrix. The method of calculation provides an alternative to the usual sum over periodic trajectories: 
it sums over paths with coincident endpoints, and includes non-vanishing boundary terms. 
An appropriately modified expansion into Matsubara modes provides a natural separation of 
the zero-mode physics. This feature may be useful in the treatment of infrared divergences 
that plague the perturbative approach in thermal field theory.
\end{abstract}


\maketitle

\section{Introduction}

Quantum statistical physics provides the conceptual and computational framework for the treatment 
of interacting many-body systems in thermal equilibrium with a heat bath. It describes a great variety 
of systems and phenomena over a wide range of scales, thus covering practically all areas of physics, 
from cosmology to particle physics, with an extensive number of examples in condensed matter.

Its broad spectrum of applications includes special relativistic systems. For those, the quantum 
statistical treatment is known as finite-temperature field theory \cite{FTFT-books}, which 
is synonymous to quantum statistical field theory. In fact, just as quantum statistical mechanics 
adds stochastic probabilities to the quantum probabilities of quantum mechanics, finite temperature 
field theory does likewise with respect to quantum field theory, the natural combination of special 
relativity and quantum mechanics.

Quantum statistical physics can be formulated in the language of euclidean (imaginary-time) 
functional integrals, quite natural for finite-temperature field theories, but also applicable in 
nonrelativistic contexts, since quantum mechanics can be viewed as a field theory in zero spatial 
dimensions. Functional integrals thus furnish a powerful unifying formalism to compute correlations.

Indeed, the formalism expresses the partition function of a given system as a generating functional 
of euclidean Green functions (correlations), similarly to zero-temperature field theory, but with the 
time direction made compact, and specific boundary conditions that constrain the domain of field 
configurations in the functional integral.

In order to best exploit the unifying aspect of the formalism, we propose to view the partition function 
as an integral of the diagonal density matrix element of the theory. The integral is performed over the 
stochastic variable which characterizes the representation of the matrix element (position representation 
for quantum mechanics, field representation for second quantized quantum mechanics or field theory). 
The density matrix is just the Boltzmann operator, whose Hamiltonian operator may either come from 
(second quantized) quantum mechanics or from field theory.

The density matrix element may be written as a functional integral in euclidean time $\tau$ which 
starts from a given configuration (a point in quantum mechanics, a field configuration in field theory) 
at $\tau=0$, and returns to that same configuration at $\tau=\beta\hbar$. Clearly, there is a sum to be 
performed over configurations which coincide at the endpoints of the euclidean time interval. Those 
boundary configurations are identified as the stochastic variable of the remaining integral. Furthermore, 
it will be shown that they are zero modes of a modified Matsubara series. The theory of the zero modes 
is the effective stochastic theory obtained from the original Hamiltonian.

This density matrix method for deriving an effective stochastic problem was already used 
in \cite{deCarvalho:2001xv}, where it led to the construction of dimensionally reduced effective 
actions. Likewise, in Ref. \cite{Bessa:2007vq} we used the density matrix method, and a semiclassical 
approximation, to investigate the thermodynamics of scalar fields. In the latter reference, the boundary 
configuration of the field (written as a constant plus gaussian fluctuations) played a very nontrivial role 
in the semiclassical formulation for the partition function of scalar fields.

Our prior uses of the method drew upon generalizations of a thermal semiclassical treatment 
previously developed and applied to quantum statistical mechanics, which produced excellent 
results for the ground state energy and the specific heat of the anharmonic (quartic) 
oscillator \cite{deCarvalho:1998mv,deCarvalho:1999fi,deCarvalho:2001vk}. In those prior uses, 
however, the connection with the stochastic variable was indirect, through classical solutions of 
equations of motion. Now, we have opted instead for a direct connection with the stochastic variable, 
to be identified with a zero mode, because we believe this is physically relevant to the discussion 
of infrared problems in several applications of quantum statistical physics. 

In the context of finite-temperature field theory, for instance, it is known that a naive implementation of 
perturbation theory for the calculation of Feynman diagrams is ill-defined in the presence of 
massless bosons. This is due to the appearance of severe infrared divergences, brought 
about by the vanishing bosonic Matsubara mode in thermal propagators. The divergences plague the 
entire series, making it essentially meaningless \cite{FTFT-books,Kraemmer:2003gd,Andersen:2004fp}.
 
As a result, one is forced to resort to resummation techniques that reorganize the perturbative series, and 
resum certain classes of diagrams, in order to extract sensible results. There are several ways 
of performing resummations, and rewriting the degrees of freedom more efficiently in terms 
of quasiparticles; we refer the reader to the reviews \cite{Kraemmer:2003gd,Andersen:2004fp}, 
and to Ref. \cite{Bessa:2007vq}, for a discussion and a list of specific references. 

All such techniques are designed to partially tame the infrared divergences, creating a non-zero domain 
of validity for weak-coupling expansions\footnote{In the case of thermal QCD, for instance, the 
domain of validity of the naive perturbative expansion is the empty set \cite{Braaten:2002wi}.}. 
Nevertheless, the zero-mode problem remains, and the region of validity of resummed perturbation 
treatments can not be indefinitely enlarged.

The infrared problems of finite temperature field theory resemble those encountered in the functional 
integral treatment of various condensed matter systems \cite{Popov}, where they are connected to 
collective excitations. This suggests that one should explore the unifying feature of the formalism, which 
is its statistical nature, to profit from parallel physical interpretations and insights. This is yet another 
reason for resorting to the density matrix formulation.

In the present article, we resume the study of quantum statistical mechanics, viewed as an exercise 
in zero-dimensional finite-temperature field theory. Our objective is to show that boundary configurations 
are the zero-mode stochastic variables of the effective statistical theory, and to compute that theory in 
various approximate schemes. Although we restrict our analysis to quantum mechanical examples, 
this should be regarded as a preliminary to the field theory case. 

We compute the partition function of an interacting one-dimensional  quantum mechanical system 
at finite temperature from a path-integral representation for the density matrix. The method of calculation 
that we propose provides an alternative to the usual sum over periodic trajectories: it sums over paths 
with coincident endpoints, and includes the contribution of non-vanishing boundary terms. Indeed, an 
appropriately modified expansion into Matsubara modes provides a natural separation of the zero-mode 
physics, which is connected to the infrared problems in thermal field theory and elsewhere, and relates 
the zero mode to the boundary value of the quantum-mechanical coordinate. 

The paper is organized as follows: in Section II, we present the proposed modified series 
expansion in Matsubara frequencies, and make a detailed comparison of our density matrix 
procedure with the standard one, for clarity; in Section III, we relate the boundary value 
of the quantum-mechanical coordinate (the stochastic variable) to the zero mode; 
In Section IV, we compute several thermodynamic quantities for the quadratic and the 
quartic potentials, and compare our results to the semiclassical findings of \cite{deCarvalho:1998mv}, 
and to some exact results; Section V contains our conclusions and outlook.

\section{A modified series expansion}\label{altMatsubara}

The partition function $Z$ for a one-dimensional quantum-mechanical system in contact with 
a thermal reservoir at temperature $T=1/(k_B\beta)$ may be written as
\begin{equation}\label{eq:Z0}
Z = \int dx_0 \,\rho \left( \beta;x_0,x_0\right )\;,
\end{equation}
where $\rho(\beta;x_0,x_0)$ is the diagonal element of the density matrix. In a path-integral 
formulation, the density matrix $\rho$ is obtained from the imaginary-time evolution of the 
trajectory $x\!:[0,\beta\hbar] \rightarrow \IR$ determined by an euclidean action $S$:
\begin{equation}
\rho(\beta;x_0,x_0) = \int\limits_{x(0)=x(\beta\hbar)=x_0} [{\cal D}x]\; e^{-S[x]/\hbar} \;,
\end{equation} 
\begin{equation}
S[x] = \int_{0}^{\beta\hbar} d\tau \, \left [ \frac{1}{2}\,m\dot{x}^2(\tau) + V(x(\tau))\right ]\;,
\end{equation}
where $V$ is the potential\footnote{In order to simplify the notation, we will adopt natural units
 where $\hbar =1$ and $\k_B=1$.}. In such a formalism, it is natural to regard $x_0$ as the effective 
stochastic degree of freedom of the theory, and to compute the distribution of the variable 
$x_0$ by integrating over the auxiliary variable $x$, with boundary conditions determined 
by $x_0$. In this article, we want to show that the effective theory for the static variable $x_0$ 
not only underscores its statistical mechanical character, but can also be identified with the 
theory of the zero mode of an appropriately modified Matsubara expansion. 

We can show that the formalism requires a modified Matsubara expansion for $x$ by calculating 
the partition function of the harmonic oscillator in two different manners. The action $S$, in this 
simple case, is given by
\begin{equation}
\label{actionQM}
S[x] = \frac{1}{2}\int_0^\beta d\tau\, \left[ m\dot{x}^2(\tau) + m\omega^2x^2(\tau)\right ] \;.
\end{equation}
Without any loss of generality, one can take $m=1$ in the previous equation by rescaling $x$ by a factor $1/\sqrt{m}$. One may compute the partition function using the following path integral:
\begin{equation}
\label{eq:Z1}
Z = \int\limits_{x(0)=x(\beta)} [{\cal D}x] \,e^{-S[x]} \;.
\end{equation}
The condition $x(0) = x(\beta)$ is usually implemented by expanding the path $x$ in 
Matsubara modes:
\begin{equation}\label{matsuba}
x(\tau) \;=\;\sum_{n=-\infty}^{\infty}c_n\,e^{-i\omega_n\tau }\;\;\;\hbox{with } \omega_n = \frac{2\pi n}{\beta}\;.
\end{equation}
The standard procedure is, then, to integrate Eq. \eqref{actionQM} by parts:
\begin{align}\nonumber
S[x] = \frac{1}{2}\int_0^\beta d\tau\, x(\tau)&\left [-\frac{d^2}{d\tau^2} + \omega^2\right ]x(\tau)\;\\\label{actionQM2}
&\;\;\;\;\;\;\;\;\;\;\;\;+ \frac{1}{2}\bigg[x(\tau)\dot{x}(\tau) \bigg ]_{0}^{\beta}\;.
\end{align}
Using the periodicity of Eq. \eqref{matsuba}, one shows that the boundary term 
in \eqref{actionQM2} is zero.
The remaining term can be cast in the form
\begin{align}\label{actionQM3}
S[x] =\frac{1}{2}\int_0^\beta d\tau d\tau^\prime\, x(\tau)\,K(\tau,\tau^\prime)\,x(\tau^\prime)\,,
\end{align}
where the Green function $K(\tau,\tau^\prime) = \Delta_F(\tau-\tau^\prime)$ is such that
\begin{subequations}
\begin{gather}
 \left[-\frac{d^2}{d\tau^2} + \omega^2 \right]\,\Delta_F(\tau)\;=\;\delta(\tau) \, ,\\
\Delta_F(\tau-\beta)= \Delta_F(\tau)\;.
\end{gather}
\end{subequations}
This corresponds to the standard free propagator \cite{FTFT-books}
\begin{align}
\Delta_F(\tau) = \frac{1}{2\omega}\left [(1+ n(\omega))~ e^{-\omega\tau} + 
n(\omega)~e^{\omega\tau}\right ]\;,
\end{align}
where $n(\omega)$ is the Bose-Einstein distribution. Using $\Delta_F$, it is straightforward 
to obtain the partition function. Up to an infinite constant, one obtains the well known result 
for the free energy
\begin{align}\label{stdZSHO}
-\frac{1}{\beta}\log Z = \frac{\omega}{2} + \frac{1}{\beta}\log (1-e^{-\beta \omega})\;.
\end{align}

There exists, however, an alternative path-integral method to calculate the partition 
function of the harmonic oscillator, which has the advantage of being {\it free of spurious 
divergences}. We start with \eqref{eq:Z0}, and decompose the configuration $x$ as
\begin{equation}
x(\tau) = x_{c}(\tau) + y(\tau)\;,
\end{equation}
where $x_{c}$ is the solution of the following Euler-Lagrange equation:
\begin{subequations}
\begin{gather}
\frac{d^2x_{c}}{d\tau^2} -\omega^2 x_c = 0 \, ,\\
x_c(0) = x_c(\beta) = x_0\;,
\end{gather}
\end{subequations}
which is given by
\begin{align}\label{classicalpath}
x_{c}(\tau) = x_0\frac{\cosh[\omega(\tau-\beta/2)]}{\cosh (\omega\beta/2)}\;.
\end{align}

It is easy to show that
\begin{align}\nonumber
S[x] &= \frac{1}{2}x_0 [\dot{x}(\beta)-\dot{x}(0)] \;+\;S[y]\\\label{dotbetadotzero}
&=\frac{x_0^2\omega (\cosh \beta \omega-1)}{\sinh \beta \omega} \;+\;S[y]\;.
\end{align}
Using that $y$ vanishes at $\tau=0,\beta$, one rewrites the action of the trajectory $y$ as
\begin{align}
S[y]
&= \frac{1}{2}\int d\tau d\tau^\prime \,y(\tau) G_y^{-1}(\tau,\tau^\prime)y(\tau^\prime) \,
\end{align} 
where the Green function $G_y$ satisfies the following equation:
\begin{subequations}\label{Gy}
\begin{gather}
\left [-\frac{d^2}{d\tau^2} + \omega^2\right ]\,G_y(\tau,\tau^\prime)\;=\;\delta(\tau-\tau^\prime) \, ,\\
G_y(0,\tau^\prime) = G_y(\beta,\tau^\prime) = 0\;.
\end{gather}
\end{subequations}
One can show \cite{deCarvalho:1998mv} that
\begin{align}\label{Gy2}
G_y(\tau,\tau^\prime) = \frac{\sinh [\omega (\tau_{>}-\beta)]
\sinh(\beta \tau_<)}{\omega \sinh(\omega \beta)}\;,
\end{align}
and
\begin{align}
\det G_y = \frac{\omega}{2\pi\,\sinh(\omega \beta)}\;.
\end{align}
In Eq. \eqref{Gy2}, $\tau_{>} = \hbox{max}(\tau,\tau^\prime)$ and  $\tau_{<} = \hbox{min}(\tau,\tau^\prime)$.
The path integral over $y$ produces, then, 
\begin{align}\label{Zcalc1}
Z&=\left(\det G_y \right)^{1/2}\, \int dx_0 
e^{-\omega\left[(\cosh \beta\omega -1)/\sinh\beta\omega\right]x_0^2}\;\\
&=(2\cosh \beta\omega -2)^{-1/2}\;.
\end{align}
Using $(2\cosh \omega\beta - 2) = [1-\exp (-\beta\omega)]^2\exp(\beta\omega)$, and
taking the logarithm we recover Eq. \eqref{stdZSHO}.

Notice, however, that the two methods treat in a different way the quantity
\begin{align}\label{qdot0b}
\dot{x}(\beta) \,-\, \dot{x}(0)\;.
\end{align}
Let us focus on the usual Matsubara expansion of the classical path (Eq. \eqref{classicalpath}). 
Being periodic, the corresponding series has a periodic derivative, so that the quantity \eqref{qdot0b} 
is put to zero. That was used to eliminate the boundary term in Eq. \eqref{actionQM2}. 
However, we see from \eqref{dotbetadotzero} that $\dot{x_c}(\beta)-\dot{x_c}(0) \neq 0$, 
otherwise one would obtain an incorrect result for $Z$. 

The second method of calculation sums over paths which have coincident endpoints, and 
thus includes paths that are {\it not} periodic. It is easy to realize that we introduce an error 
in the value of the action of any non-periodic path $x$ by evaluating the action of the 
associate Matsubara series as given by Eq. \eqref{matsuba}. The reason is that the usual Matsubara 
expansion is not suitable to handle boundary terms involving the derivative of $x$ at $0$ and 
$\beta$. We stress that, as long as the only condition over paths in the path integral is to have 
coincident values at $0$ and $\beta$, non-periodicity {\it is} allowed.

\begin{figure}[t!]
\begin{center}
\resizebox*{!}{6.0cm}{\includegraphics{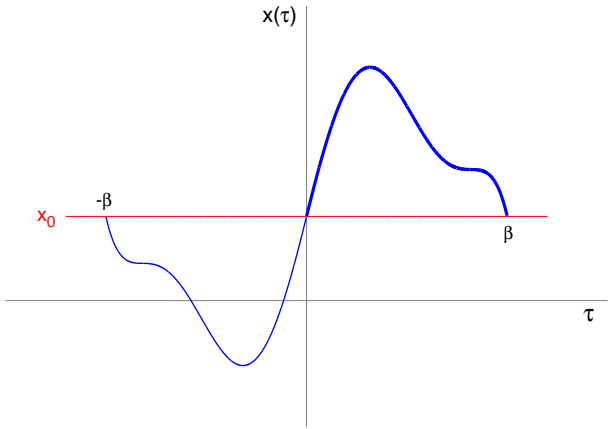}}
\end{center}
\vspace{-5mm}
\caption{\label{oddext}Schematic construction of the extension of a configuration $x$ defined 
originally in $[0,\beta]$ to the interval $[-\beta,\beta]$. Notice the odd character of the extension 
of $x-x_0$.}
\end{figure}

One can remedy this situation by defining an extension of $x$ to $[-\beta,\beta]$ as 
follows (see Fig. \ref{oddext}):
\begin{equation}\label{phi_extensionQM}
\tilde{x}(\tau) = x(\tau)\,\chi_{[0,\beta]} + (2x_0 - x(-\tau))\,\chi_{[-\beta,0)}\;,
\end{equation}
where $\chi_{A}$ denotes the characteristic function of $A \subset \IR$. Notice that 
$\tilde{x}(-\tau)-x_0 = -(\tilde{x}(\tau) -x_0)$. The odd character of the extension
 implies that $\tilde{x} - x_0$ has a series of sines:
\begin{equation}\label{seriesfortildeq}
\tilde{x}(\tau) - x_0 \,\fs\, \sum_{n=1}^{\infty}\tilde{x}_n\,\sin \hat{\omega}_n \tau\,, \;\;\hbox{with } \hat{\omega}_n = \frac{n \pi}{\beta}\;.
\end{equation}
The notation with a dot is a reminder that the r.h.s. of 
Eq. \eqref{seriesfortildeq} is a Fourier approximation of the l.h.s. The desired restriction to 
$[0,\beta]$ is:
\begin{equation}\label{varphizeromode}
x(\tau) \fs x_0 + \sum_{n=1}^{\infty}x_n\,\sin \hat{\omega}_n \tau\,,
\end{equation}
where
\begin{equation}
x_n = \frac{2}{\beta}\,\int_{0}^{\beta}d\tau\, x(\tau)\,\sin \hat{\omega}_n\tau\;.
\end{equation}
The series \eqref{varphizeromode} does not impose conditions on the derivative of $x$. 
It is easy to convince oneself that evaluating the euclidean action $S$ with the r.h.s. of 
Eq. \eqref{varphizeromode} one obtains exactly $S[x]$. In particular, we have for the 
classical solution:
\begin{align}
x_{c}(\tau)\, \fs \,x_0-x_0\,\sum_{n \;\tiny\hbox{odd}}\frac{4\omega^2}{n \pi\,(\hat{\omega}_n^2  
+  \omega^2)}\, \sin \hat{\omega}_n\tau\;,
\end{align}
so that the quantity $\dot{x_c}(\beta)-\dot{x_c}(0)$ is given by
\begin{align}
\sum_{n \;\tiny\hbox{odd}}\frac{-4x_0\,\omega^2\,\hat{\omega}_n}{n \pi\,(\hat{\omega}_n^2  
+  \omega^2)}\,\left( \cos \hat{\omega}_n\beta -1 \right ) = 
\frac{2x_0\omega(\cosh \beta \omega -1)}{\sinh \beta \omega}\;,
\end{align}
as implied by Eq. \eqref{dotbetadotzero}. Obviously, the series defined by Eq. \eqref{varphizeromode} 
does not correspond to the most general trajectory in $Z$. In particular, that series imposes a serious 
restriction on the second derivative of $x$, by forcing it to vanish at $0$ and $\beta$. However, at 
least for the calculation of usual action functionals, the boundary terms do not involve second 
derivatives and the aforementioned restriction is actually unimportant.
%
%
\section{The boundary value as the zero mode}
The first remarkable property of the decomposition \eqref{varphizeromode} is that the boundary 
value $x_0$ is the zero, static component of $x$ which, in the context of thermal field theories, 
plays a special role in the infrared (low-momentum) limit of the theory. As mentioned previously, 
the infrared physics of bosonic theories is not accessible to plain perturbation theory, and 
resummation techniques are required in order to produce sensible results. In common, all such 
techniques render a special treatment to the zero mode\footnote{Rigorously speaking, $x_0$ is 
not a true Fourier mode since it is not orthogonal to $\sin \hat{\omega}_n\tau$ in $[0,\beta]$, but in 
$[-\beta,\beta]$. With this caveat in mind, we will continue to refer to $x_0$ as a mode of $x$.}.
 
It is important to stress that the zero mode $x_0$ is {\it not} the zero mode of the usual Matsubara 
expansion. In fact, from Eq. \eqref{matsuba}, one can notice that $x_0$ involves the coefficients of 
all Matsubara modes:
\begin{equation}
x_0 = \sum_{n=-\infty}^{\infty} c_n\;.
\end{equation}

The present identification of the zero mode with the boundary field in the 
context of quantum statistical mechanics aggregates to the zero-mode physics 
yet another important feature: the connection with the classical behavior at high temperatures. 
That regime is characterized by the condition
\begin{equation}
\hbar \omega \ll k_B T
\end{equation}
(or $\beta \omega \ll 1$, in natural units), where $\hbar \omega$ is the typical spacing between 
quantum levels. In order to see why the static mode governs the classical limit of the partition 
function, let us consider $x_c$, the solution of the equation of motion: 
\begin{subequations}\label{eulerlagrange}
\begin{gather}
\frac{d^2x_c}{d\tau^2} - V'(x_c(\tau)) = 0 \, ,\\
x_c(0) = x_c(\beta) = x_0\;.
\end{gather}
\end{subequations}
In the high-temperature regime ($\beta\omega \ll 1$), thermal fluctuations dominate. In this limit, 
$x_c \rightarrow x_0$ and $S(x_c) \rightarrow \beta V(x_0)$, and we obtain for the partition function
\begin{equation}
\lim_{T \rightarrow\infty} Z \;=\; N\;\int dx_0 \, e^{-\beta V(x_0)}\;,
\end{equation}
where $N$ is a normalization factor which incorporates quantum fluctuations. A proper 
normalization of $Z$ leads to $N^2=m\,k_BT/(2\pi\hbar^2)$, rendering the semiclassical expression
\begin{equation}\label{classicalZ}
\lim_{T \rightarrow\infty} Z \;=\; \int \frac{dp}{2\pi\hbar}\,\int dx_0 \,\hbox{e}^{-H/k_B T}\,,
\end{equation}
with $H = p^2/2m + V(x_0)$, $m$ being the particle mass. 

As discussed in \cite{Kleinert:2004ev}, for $\beta \omega \ll 1$ the correlation 
$\langle x^2(\tau)\rangle$  follows the classical linear scaling with $T$. This behavior is entirely 
associated with the static component of $x$, and represents a problem for the high-temperature 
limit of the usual perturbative expansion. In contrast, the subtracted thermal propagator (without 
the contribution from the static mode) goes to zero as $T$ increases, so that one should strongly 
improve the convergence of the perturbation expansion by calculating diagrams with that subtracted 
propagator.

We conclude that, either to control infrared divergences in thermal field theories or to describe 
the classical limit in quantum statistical mechanics, we need a separate treatment of the zero mode. 
The main advantage of our approach is to provide this separation in a natural fashion.

In order to explore the zero-mode physics, it is convenient to write
\begin{equation}\label{q=q0+eta}
x(\tau)\;=\; x_0 + y(\tau)\,,
\end{equation}
where $y$ corresponds to the sinusoidal modes of Eq. \eqref{varphizeromode} and 
has the following property:
\begin{equation}
y(0) =y(\beta)=0\;.
\end{equation}
In terms of $y$ one can express $Z$ conveniently as
\begin{equation}\label{eq:Z1}
Z = \int dx_0\,\int \limits_{y(0)=y(\beta)=0} [{\cal D}y] \, e^{-S[x_0+y]}\;,
\end{equation}
and different strategies can be used to handle $S[x_0 + y]$ in order to obtain an effective 
theory\footnote{With the normalization factor used in \eqref{eq:Zeff}, one has the 
property: $V_{\tiny \hbox{eff}}(x_0) \rightarrow V(x_0)$ when $T \rightarrow \infty$, as 
follows from \eqref{classicalZ}.} for $x_0$:
\begin{equation}\label{eq:Zeff}
Z = \int \frac{dx_0}{\sqrt{2\pi \hbar^2/(m\,k_BT)}}\,e^{-\beta\, V_{\tiny \hbox{eff}}(x_0)}\;.
\end{equation}
Notice that the intermediate step, including the paths $y$, is built using quantities which vanish 
at the boundary of imaginary time, and the contribution from problematic static quantities is factorized 
in the final integration over $x_0$. Besides, for interacting systems, the interaction potential evaluated
 at $x(\tau)  = x_0 + y(\tau)$ can produce a $y$-dependent quadratic term for the zero mode $x_0$, which
could prevent possible divergences in the effective theory.

In the sequel, we use a very simple expansion around $x_0$ in two cases: the harmonic potential, 
as a consistency check, and the single-well quartic potential, in order to test the quality of our
alternative approach and compare it to previous descriptions.

%
\section{Applications and comparison to previous results}

\subsection{The quadratic case revisited}

The action is again the one given by Eq. \eqref{actionQM}. Now, we use the decomposition 
of $x$ given by Eq. \eqref{q=q0+eta} to obtain
\begin{align}\nonumber
S[x]
&= \frac{\beta\omega^2}{2}x_0^2 + \frac{1}{2}\int d\tau d\tau^{\prime}\left \{y\,G_{y}^{-1}\,y 
+ 2\omega^2 x_0 y \right \} \\\nonumber
&= \frac{\beta\omega^2}{2}x_0^2 + \frac{1}{2}\int d\tau \big \{ \left( y 
+ \omega^2x_0 \,G_{y}\right ) \,G_{y}^{-1}\,\times\\
&\;\;\;\;\;\;\;\; \left( y + \omega^2x_0 \,G_{y}\right )\big \} 
-\frac{\omega^4x_0^2}{2}\,\int d\tau d\tau^{\prime} G_y\;,
\end{align}
where $G_y$ was defined in \eqref{Gy}. A simple calculation gives
\begin{align}\label{intintG}
\int  d\tau d\tau^\prime G_y(\tau,\tau^\prime) = \frac{\beta}{\omega^2}  
- \frac{2(\cosh \beta \omega -1)}{\omega^3\sinh \beta \omega}\,.
\end{align}
Shifting the variable of integration to 
\begin{equation}
\zeta(\tau) =  y(\tau) + \omega^2x_0 \, \int d\tau^\prime\, G_{y}(\tau,\tau^\prime)\;,
\end{equation}
which obeys $\zeta(0)=\zeta(\beta)=0$, and performing the gaussian integration, we obtain
\begin{align}
Z 
&=\left(\det G_y \right)^{1/2}\, \int dx_0 
e^{-\omega\left(\cosh \beta\omega -1)/\sinh\beta\omega\right)x_0^2}\;.
\end{align}
Integrating over $x_0$ and taking the logarithm, one reproduces Eq. \eqref{Zcalc1}. Notice that 
the propagator used in the perturbation expansion of the integral over the path $y$ is well-behaved 
in the limit of small $\omega$:
\begin{align}\label{highT}
| G_y(\tau,\tau^\prime) | \;\leq\; \frac{\tanh \beta \omega/2}{2\omega}\; \stackrel{\beta \omega \ll 1}{\longrightarrow} \;\frac{\beta}{4}\;.
\end{align}
%

\subsection{The anharmonic oscillator}

Let us now consider the following action:
\begin{equation}\label{actionQM4}
S[x] = \int_0^\beta\!\! d\tau \left [\frac{1}{2}m\dot{x}^2(\tau) + \frac{1}{2}m\omega^2 x^2(\tau) 
+ \frac{\lambda}{4}x^4(\tau) \right]\;.
\end{equation}
It is convenient to define the dimensionless variables: $q=\sqrt{\lambda/(m\omega^2)}\,x$,  
$\theta = \omega\tau$, $\Theta = \omega\beta$ and $g=\lambda/(m^2\omega^3)$. The associated 
action for the variable $q$ is:
\begin{equation}\label{actionQM5}
{\cal S}[q] = \frac{1}{g}\,\int_0^{\Theta}\!\! d\theta \left (\frac{1}{2}\dot{q}^2(\theta) 
+ \frac{1}{2}q^2(\theta) + \frac{1}{4}q^4(\theta)\right)\;.
\end{equation}
Using the decomposition $q(\theta) = q_0 + \sqrt{g}\,\eta(\theta)$, we write for the action:
\begin{equation}
{\cal S}[q] = {\cal S}_2[q_0,\eta] + {\cal S}_I[q_0,\eta]\;,
\end{equation}
where ${\cal S}_2$ contains ${\cal S}[q_0]$, the kinetic term, and those which are linear or
quadratic in $\eta$. The remaining contributions to ${\cal S}[q]$ are collected in ${\cal S}_I$. 
It is a simple matter to show that
\begin{align}\nonumber
{\cal S}_2[q_0,\eta] = {\cal S}[q_0]+\frac{1}{2}\int_0^\Theta d\theta \,\bigg \{\eta(\theta) &\left [-\frac{d^2}{d\theta^2} 
+ \ow^2 \right]\eta(\theta)\\
&\;\;\;\; +\frac{2\alpha}{\sqrt{g}}\, \eta(\theta)\bigg\}\,,
\end{align}
and
\begin{equation}\label{SI(q)}
{\cal S}_I[q_0,\eta] \,=\, \int_0^\Theta d\theta\,\left [\sqrt{g}\,q_0 \eta^3(\theta) + 
\frac{g}{4}\eta^4(\theta)\right ]\;,
\end{equation}
where $\ow^2  = 3\, q_0^2 + 1$ and $\alpha =  q_0^3 + q_0$. One obtains a quadratic 
approximation for the partition function by considering
\begin{equation}\label{eq:Z_2}
Z_2 = \int dq_0\,\int \limits_{\eta(0)=\eta(\Theta)=0} [{\cal D}\eta] \, e^{-{\cal S}_2[q_0,\eta]}\;.
\end{equation}

Proceeding as in the previous section, we perform the gaussian integration over the variable
\begin{equation}
\zeta(\theta) =  \eta(\theta) + \frac{\alpha}{\sqrt{g}}\,\int_0^\Theta 
d\theta^\prime\, G_\eta(\theta,\theta^\prime)\;,
\end{equation}
where the Green function $G_\eta$, defined by
\begin{subequations}\label{Geta}
\begin{gather}
\left [-\frac{d^2}{d\theta^2} + \ow^2\right ]\,G_\eta(\theta,\theta^\prime)\;=\;\delta(\theta-\theta^\prime)\, ,\\
G_\eta(0,\theta^\prime) = G_\eta(\Theta,\theta^\prime) = 0\;,
\end{gather}
\end{subequations}
is given by
\begin{align}
G_\eta(\theta,\theta^{\prime}) = \frac{\sinh [\ow (\theta_{>}-\Theta)]
\sinh(\ow\, \theta_<)}{\ow \sinh(\ow\, \Theta)}\;.
\end{align}
The result for $Z_2$ is then
\begin{align}\label{eq:Z_2b}
Z_2 = \int  dq_0& \left ( \hbox{det }G_\eta\right )^{1/2}\exp\{-{\cal S}[q_0] + \Sigma_\eta\}\,,
\end{align}
where
\begin{align}
\det G_\eta = \frac{\ow}{2\pi\,\sinh(\ow\, \Theta)}\;
\end{align}
and
\begin{align}
\Sigma_\eta = \,\frac{\alpha^2}{2g}\int_0^\Theta d\theta d\theta^\prime G_\eta (\theta,\theta^\prime)\;.
\end{align}
To go beyond the quadratic order in $\eta$, we use the following approximation:
\begin{equation}\label{correctionlambda}
e^{-{\cal S}_I[q_0,\eta]} \,\approx\, 1 - \int_0^\Theta d\theta\,\left [\sqrt{g}\,q_0 \eta^3(\theta) + 
\frac{g}{4}\eta^4(\theta)\right ]\;. 
\end{equation}
Therefore, $Z \approx Z_2 + \delta^{(1)}Z$, where
\begin{align}\label{deltaZ}
\delta^{(1)}Z = -\int  dq_0 
\,\int_0^\Theta d\theta\,\left [\sqrt{g}\,q_0 \langle\eta^3(\theta)\rangle\, + 
\frac{g}{4}\langle \eta^4(\theta)\rangle \,\right ]\;,
\end{align}
with
\begin{align}\label{correlations}
\langle \eta(\theta_1)\ldots\eta(\theta_k)\rangle \, =\!\!\!\!\! 
\int \limits_{\eta(0)=\eta(\Theta)=0}\!\!\!\!\! [{\cal D}\eta] \, 
e^{-{\cal S}_2[q_0,\eta]}\eta(\theta_1)\ldots\eta(\theta_k)\;.
\end{align}
As usual, we can introduce a linear coupling of $\eta$ with an external current $j$:
\begin{equation}\label{eq:Z_2[j]}
Z_2[j] = \int dq_0\!\!\!\!\!\!\!\!\!\!\int \limits_{\eta(0)=\eta(\Theta)=0} \!\!\!\!\!\![{\cal D}\eta] 
\exp \left\{-{\cal S}_2[q_0,\eta] + \int d\theta j(\theta)\eta(\theta)\right \},
\end{equation}
and obtain the correlations \eqref{correlations} as functional derivatives of $Z_2[j]$ with respect to $j$.
With minor  modifications on the recipe for $Z_2$, one obtains
\begin{align}\label{eq:Z_2[j]b}
  Z_2[j] = \int  dq_0& \left ( \hbox{det }G_\eta\right )^{1/2}\exp\{-{\cal S}[q_0] + \Sigma_\eta[j]\}\;,
\end{align}
where
\begin{align}
\Sigma_\eta[j] = \frac{1}{2}\int_0^\Theta d\theta d\theta^\prime \sigma(\theta)\og_\eta (\theta,\theta^\prime)\sigma(\theta^\prime)\;,
\end{align}
and $\sigma(\theta) = \alpha/\sqrt{g} - j(\theta)$. Finally, we use
\begin{align}\label{d3dj3}
\hbox{e}^{-\Sigma_\eta}\,\left [\frac{\delta^3}{\delta j^3(\theta)} \right ]_{j=0} \hbox{e}^{\;\Sigma_\eta[j]}\,= \,I_0^3(\theta)+ 3\,I_0(\theta)\,G_\eta(\theta,\theta)\,
\end{align}
and
\begin{align}\nonumber
\hbox{e}^{-\Sigma_\eta}\,\left [\frac{\delta^4}{\delta j^4(\theta)} \right ]_{j=0}\!\!\!\! \hbox{e}^{\;\Sigma_\eta[j]} = I_0^4(\theta)+& 6\,I_0^2(\theta)\,G_{\eta}(\theta,\theta)\\\label{d4dj4}
&+3G_{\eta}^2(\theta,\theta)\,,
\end{align}
where $I_0(\theta)$ plays the role of an expectation value of $\eta(\theta)$:
\begin{align}\nonumber 
I_0(\theta) &= -\frac{\alpha}{\sqrt{g}}\,\int_0^\Theta d\theta^\prime \,G_\eta(\theta,\theta^\prime)\, \\
&=\frac{\alpha}{\sqrt{g}}\,\frac{\cosh \ow \theta - \sinh \ow\theta \,\tanh \ow\theta/2 -1}{\ow^2}.
\end{align}
The remaining integrations over $\theta$ can be done analytically, providing all the ingredients 
for the calculation of the first correction to $Z_2$. 

\begin{figure}[t]
\begin{center}
\resizebox*{!}{5.5cm}{\rotatebox{270}{\includegraphics{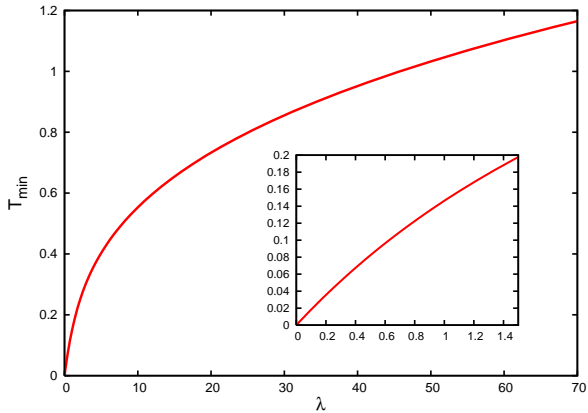}}}
\end{center}
\vspace{-5mm}
\caption{\label{fig:delta1} Plot of $T_{\tiny\hbox{min}}$ as a function of the coupling constant $\lambda$ ($m=1,\omega=1$). 
The quadratic approximation is applicable for temperatures larger than $T_{\tiny \hbox{min}}$.}
\end{figure}

Notice that the integrand of \eqref{eq:Z_2b} is dominated by the vicinity of $q_0=0$, where the 
quadratic action reaches its minimum value. For small temperatures, that integrand becomes 
sharply peaked at zero, while for large $T$ a broad range of $q_0$ substantially contributes. 
Therefore, one can estimate the importance of the correction term \eqref{deltaZ} using the 
following quantity:
\begin{align}
\langle {\cal S}_I(0,\eta)\rangle \, = \frac{3g}{4} \int_0^\Theta 
d\theta\,G_\eta^2(\theta,\theta)\;\;\stackrel{\Theta \gg 1}{\longrightarrow}\; \frac{3 g\,\Theta}{16}\;.
\end{align}
The temperature $T_{\tiny \hbox{min}}$ where $\langle {\cal S}_I(0,\eta)\rangle\,$ goes to $1$ is displayed 
in Fig. \ref{fig:delta1} as a function of the coupling constant $\lambda$. $T_{\tiny \hbox{min}}$ provides 
a rough estimate for the validity of the quadratic approximation. From Fig. \ref{fig:delta1}, we see that 
$T_{\tiny\hbox{min}}$ is of order $1$ for $\lambda$ as large as $50$.

\begin{figure}[t]
\begin{center}
\resizebox*{!}{6.2cm}{\includegraphics{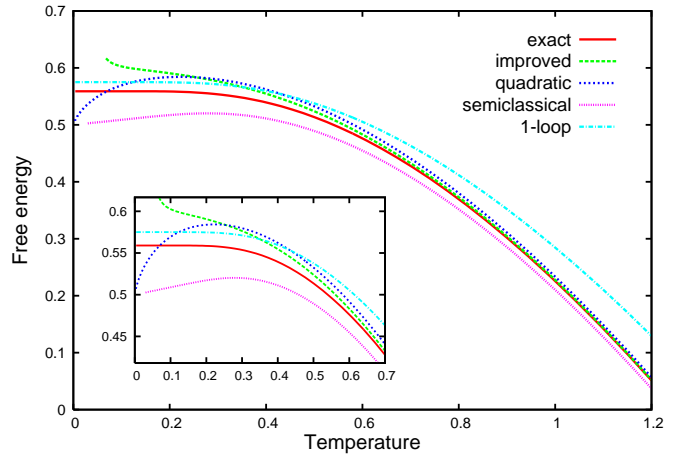}}
\end{center}
\vspace{-5mm}
\caption{\label{freeE} Free energy of the anharmonic oscillator as a function of the 
temperature for $\lambda = 0.4$, $m=1$  and $\omega=1$ (see text for details).}
\end{figure}

\begin{figure}[t]
\begin{center}
\resizebox*{!}{6cm}{\includegraphics{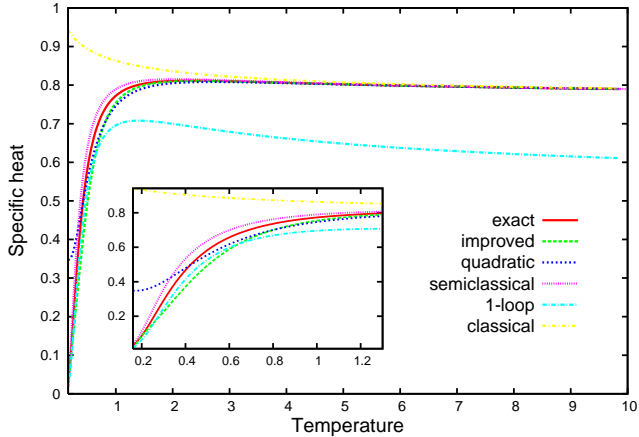}}
\end{center}
\vspace{-5mm}
\caption{\label{specheat} Specific heat of the anharmonic oscillator as a 
function of the temperature for $\lambda = 0.4$, $m=1$ and $\omega =1$ (see text for details).}
\end{figure}

The results for the free energy and for the specific heat are displayed in Fig. \ref{freeE} and 
Fig. \ref{specheat}, respectively. In the figures, {\it classical} corresponds
to calculation using Eq. \eqref{classicalZ}, {\it quadratic} uses Eq. \eqref{eq:Z_2b}, 
{\it improved} includes the correction corresponding to Eq. \eqref{deltaZ}, 
 {\it semiclassical} refers to the semiclassical result obtained in Ref. \cite{deCarvalho:1998mv}, 
{\it 1-loop} stands for the usual 1-loop perturbative result,
and {\it exact} corresponds to results from Refs. \cite{biswas_etal,hioe_montroll} combined with 
WKB estimates. 

The first remarkable thing in the plots is the failure of the perturbative result in the high-temperature 
region ($T \gg m$). As seen in Ref. \cite{deCarvalho:1998mv}, that region is well described by the 
semiclassical curve, and asymptotically, by the classical one. This is evident from both plots. It is 
interesting that, for $g=0.4$, the agreement extends to temperatures down to $T \approx m/2$,  a 
region where the character of the system is far from classical. However, it is surprising that our simple 
calculation using \eqref{eq:Z_2b} works as well as the semiclassical one in that region. 
 
Indeed, the cited semiclassical calculation is based on a quadratic expansion around exact solutions 
of \eqref{eulerlagrange} for the quartic potential. In practice, one has to deal with rather involved 
Jacobi elliptic functions \cite{deCarvalho:1998mv,AS}. 
In contrast, using the expansion \eqref{q=q0+eta} no knowledge of classical solutions is demanded. 
This simplification represents an economic alternative which can be crucial in contexts where exact 
classical solutions are not  available. From the plots, one also concludes that, in the region where the 
approximation is supposed to be good, the improved calculation exhibits a stronger convergence, 
indicating that the approximation is consistent.

A detailed look at the quadratic curve in the free energy plot (Fig. \ref{freeE}) reveals a non-monotonic 
behavior for low temperatures: the free energy passes by a maximum and decreases towards the free 
value $0.5$. This is a general feature regardless of the value of $m$ and $\lambda$, and it is a clear signal 
of the breakdown of the approximation below $T_{\tiny \hbox{min}}$ (see Fig. \ref{fig:delta1}). The 
improved free energy diverges at $T_{\tiny \hbox{min}}$, because at this value the sum 
$Z_2 + \delta^{(1)}Z$ vanishes. The exact curve for the free energy reaches its maximum value at the 
ground state energy, where it rests down to $T=0$. Using the maximum value assumed by the quadratic 
curve as an approximate value for the ground state energy, we obtain unexpectedly reasonable results, 
as shown in Table \ref{tableGSE}.

\begin{table}[b!]
\centering
\begin{tabular}{c|c|c|c}\hline\hline
$\lambda$\; & $E_0$ (exact)
& $E_0$ (quadratic) & Error($\%$) \\
\hline
0.008 & 0.501 & 0.505 & 0.8\\
0.04 & 0.507 & 0.518 & 2.2\\
0.4\; & 0.559 & 0.584 & 4.5 \\
1.2\; & 0.638 & 0.662 & 3.8 \\
2.0\; & 0.696 & 0.718 & 3.2 \\
4.0\; & 0.804 & 0.818 & 1.7 \\
8.0\; & 0.952 & 0.958 & 0.6 \\
200.0 & 2.500 & 2.450 & 2.0 \\\hline\hline
\end{tabular}
\caption{\label{tableGSE}Ground state energy of the quartic oscillator for different values of the 
coupling $\lambda$ ($m=\omega=1$). Exact data were extracted from Ref. \cite{hioe_montroll}, 
whereas $E_0$ (quadratic) was obtained using Eq. \eqref{eq:Z_2b}. }
\end{table}

Fig. \ref{specheat} compares different predictions for the specific heat for $\lambda = 0.4$, $m=1$ 
and $\omega=1$. Notice that all curves, except for the perturbative one, have the correct high-temperature 
limit. The anomalous behavior of the quadratic approximation near $T_{\tiny \hbox{min}}$ is even more 
evident in this plot. The improvement obtained in the low-temperature regime when one 
corrects the quadratic calculation with \eqref{deltaZ} is also clearly shown.

\begin{figure}[t]
\begin{center}
\resizebox*{!}{5.5cm}{\rotatebox{270}{\includegraphics{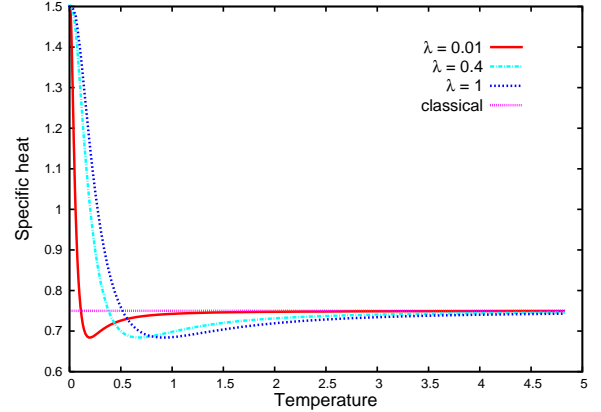}}}
\end{center}
\caption{\label{zeromass} Specific heat of the anharmonic oscillator with $\omega=0$, $m=1$ and different values of $\lambda$.}
\end{figure}

Finally, Fig. \ref{zeromass} displays the quadratic result for the specific heat when $\lambda = 0.4$ 
and $m=1$ in the limit of zero $\omega$. For large temperatures, the curve converges to the classical 
value $0.75$. As discussed in Section \ref{altMatsubara}, with the separation of the zero mode, the 
finiteness of the theory for the $\eta$ variable is not affected in the infrared limit, and we see in this 
specific case that the interaction makes the effective zero-mode theory well defined.

%

\section{Conclusions and outlook}

The theoretical description of the thermodynamics of a given interacting many-body system 
kept in thermal equilibrium by the contact with a heat bath is an issue of capital importance 
in most realms of physics, from cosmology to condensed matter settings. A very convenient 
framework is provided by finite-temperature field theory, where one can use the imaginary-time 
formalism and make a direct connection to the powerful formulation of thermal averages in terms of 
functional integrals. Nevertheless, whenever collective (massless) bosonic modes are relevant, 
as is the case in practically any plasma, infrared divergences are unavoidable in a plain 
perturbative treatment, forcing a reorganization of the series that grants special attention to 
the zeroth Matsubara mode. As discussed in this paper, we also need a separate treatment of 
the static mode already in the context of quantum statistical mechanics in order to improve 
the convergence of the perturbative expansion in the high-temperature limit.

The choice of the traditional Matsubara mode expansion, combined with the trace requirement 
that the paths (configurations) coincide at the two boundaries of the compact euclidean time, implies 
a restriction to periodic paths in the process of functional integration to compute 
the partition function. However, rigorously speaking, the paths need not be 
periodic in the path-integral representation of the partition function. Indeed, the
 necessity of paths which are not periodic is particularly clear in the formulation
 of the partition function as an integral of the diagonal density matrix element of the theory. We
 have shown that, in order to allow for non-periodicity, one has to define a modified expansion
 into Matsubara modes.

In this paper, we have followed our alternative strategy in the case of quantum statistical mechanics, viewed 
not only as a toy model for the case of finite-temperature field theory, but also as a prototype 
for various relevant systems in statistical physics. More specifically, we have explored an 
alternative way of computing the partition function in the path-integral formalism that includes 
non-periodic trajectories, for which the usual Matsubara expansion leads to an incorrect result 
for the action of the trajectory. As was shown above, one can properly incorporate the contribution 
of non-periodic paths by using a modified Matsubara series expansion in which the zero mode is 
identified with the boundary value of the path. The latter turns out to be the stochastic variable 
that survives in the final effective theory. 

The approach proposed in this paper has, thus, the advantage of providing a natural separation of the 
physics of the zero mode. In fact, we built a very simple effective theory for the zero mode in the 
nontrivial problem of the anharmonic oscillator, obtaining very precise results for practically the 
whole range of temperatures, and in a framework that is free of spurious divergences. In particular, 
we were able to describe the semiclassical (high-temperature) limit of the anharmonic oscillator without 
any information about the classical solutions. We expect that other potentials can profit from the simplicity
 of the present method in quantum mechanical systems. 

In the case of finite-temperature field theory there are, of course, subtle issues of renormalization, 
which are absent in the quantum mechanical setting, that will have to be addressed. Nevertheless, 
the results obtained in quantum statistical mechanics are encouraging, and we hope that this 
new perspective may shed light onto the problem of infrared divergences. Results in this direction 
will be presented elsewhere \cite{future}


\section*{Acknowledgments}

A.B. thanks F.~T.~C. Brandt, C.~Farina, J.~Frenkel, Ph.~Mota and A.~R. da Silva for fruitful discussions.
This work was partially supported by CAPES, CNPq, FAPERJ, FAPESP, and FUJB/UFRJ. 


\end{document}